\documentclass[print]{aa} 

\usepackage{graphicx}
\usepackage{txfonts}
\usepackage{multirow}
\usepackage{colortbl}
\usepackage{hyperref}  
\hypersetup{colorlinks=true,linkcolor=[rgb]{1.,0.2,0.2},citecolor=[rgb]{0.1,0.4,1.},filecolor=[rgb]{0.7,0.2,0.2},urlcolor=[rgb]{0.7,0.2,0.2}}

\usepackage{color}
\definecolor{blue}{rgb}{0., 0., 1}
\definecolor{lightblue}{rgb}{0.1,0.4,1.}

\usepackage{soul}

\def\HI{H\,{\sc i}}

\begin{document} 

\title{Evidence of ram-pressure stripping of WLM, a dwarf galaxy far away from any large host galaxy} 

\titlerunning{Evidence of ram-pressure stripping of WLM}
\authorrunning{Y.\ Yang et al.} 


\author{Yanbin Yang\inst{\ref{ins1}} \fnmsep\thanks{E-mail: \href{mailto:yanbin.yang@obspm.fr}{yanbin.yang@obspm.fr}}
\and Roger Ianjamasimanana\inst{\ref{ins2},\ref{ins3},\ref{ins4}}
\and Francois Hammer\inst{\ref{ins1}}
\and Clare Higgs\inst{\ref{ins5}} 
\and Brenda Namumba\inst{\ref{ins3}} 
\and Claude Carignan\inst{\ref{ins6},\ref{ins7},\ref{ins8}}
\and Gyula I. G. J\'ozsa,\inst{\ref{ins9},\ref{ins3}}
\and Alan W. McConnachie\inst{\ref{ins10}}
}
\institute{ GEPI, Observatoire de Paris, Universite PSL, CNRS, Place Jules Janssen 92195, Meudon, France \label{ins1}
\and Instituto de Astrofísica de Andalucía (CSIC), Glorieta de la Astronomía, E-18008 Granada, Spain  \label{ins2}
\and Department of Physics and Electronics, Rhodes University, PO Box 94, Makhanda, 6140, South Africa \label{ins3}
\and South African Radio Astronomy Observatory, 2 Fir Street, Black River Park, Observatory, Cape Town,  7925, South Africa \label{ins4}
\and Physics \& Astronomy Department, University of Victoria, 3800 Finnerty Rd,  Victoria, B.C., Canada, V8P 5C2 \label{ins5}
\and Department of Astronomy, University of Cape Town, Private Bag X3, Rondebosch 7701, South Africa \label{ins6}
\and D\'{e}partement de physique, Universit\'{e} de Montr\'{e}al,  Complexe des sciences MIL, 1375 Avenue Th\'{e}r\`{e}se-Lavoie-Roux Montr\'{e}al, Qc, Canada H2V 0B3 \label{ins7}
\and Laboratoire de Physique et de Chimie de l'Environnement, Observatoire d'Astrophysique de l'Universit\'{e}  Ouaga I Pr Joseph Ki-Zerbo (ODAUO), BP 7021, Ouaga 03, Burkina Faso \label{ins8}
\and Max-Planck-Institut f\"ur Radioastronomie, Radioobservatorium Effelsberg,Max-Planck-Stra{\ss}e 28, 53902 Bad M\"unstereifel, Germany \label{ins9}
\and NRC Herzberg Astronomy and Astrophysics, 5071 West Saanich Road, Victoria, B.C., Canada, V9E 2E7 \label{ins10}
}

\date{Received; accepted}


\abstract{
Dwarf galaxies are affected by all the evolutionary processes normally at work in galaxies of any mass. As fainter and less massive galaxies, however, dwarf galaxies are particularly susceptible to environmental mechanisms that can more easily perturb these systems. Importantly, the presence of nearby large galaxies is expected to have a profound effect on dwarf galaxies. Gravitational (especially tidally induced) effects from the large galaxy can cause mass to be lost from the dwarf, and the passage of the dwarf through the gaseous medium surrounding the large galaxy can additionally cause the dwarf to lose its own gas through a process called ram-pressure stripping. Such effects are considered to be the main sources of difference between ``satellite'' and ``field'' dwarf galaxy populations. Here, we report on new observations of the gaseous content of Wolf-Lundmark-Melotte (WLM), an archetype of isolated, gas-rich field dwarf galaxies in the Local Universe, which presents a much more complex situation. Previous studies of its gaseous disk suggest it has perturbed kinematics;
here, we identify four trailing, extended gas clouds lying in the direction opposite to WLM's spatial motion, as well as a spatial offset between the WLM gas and stars. Overall, the morphology and kinematics of this gas show that WLM is undergoing ram-pressure stripping, despite being 930 and 830 kpc from the Milky Way and M31, respectively. 
This finding indicates the presence of an intergalactic, gaseous reservoir far from large galaxies whose evolutionary role in galaxies, both large and small, may not be fully appreciated.
}


\maketitle

\section{Introduction}
Wolf-Lundmark-Melotte (WLM) was discovered in 1909 by Max Wolf and then confirmed by Knut Lundmark and Philibert Melotte. It is a nearby example of a field dwarf galaxy and is located near the edge of the Local Group.
This dwarf is among the smallest galaxies with a rotational field established 
in Spitzer Photometry and Accurate Rotation Curves \citep[SPARC;][]{Lelli2016}. 
Given its isolated nature and highly inclined orientation, the dark matter mass of this dwarf is considered to be robustly determined and shows that WLM's dark matter may exceed its baryonic content by a factor close to 90 \citep{Read2017}. However, its approaching and receding sides show different rotational velocities \citep{Kepley2007}, and the central region shows an unexpected minimum in velocity dispersion \citep{Ianjamasimanana2020}, which led us to suspect it has perturbed kinematics \citep{Khademi2021}.

\begin{figure*}
\centering
\includegraphics[width=0.85\textwidth]{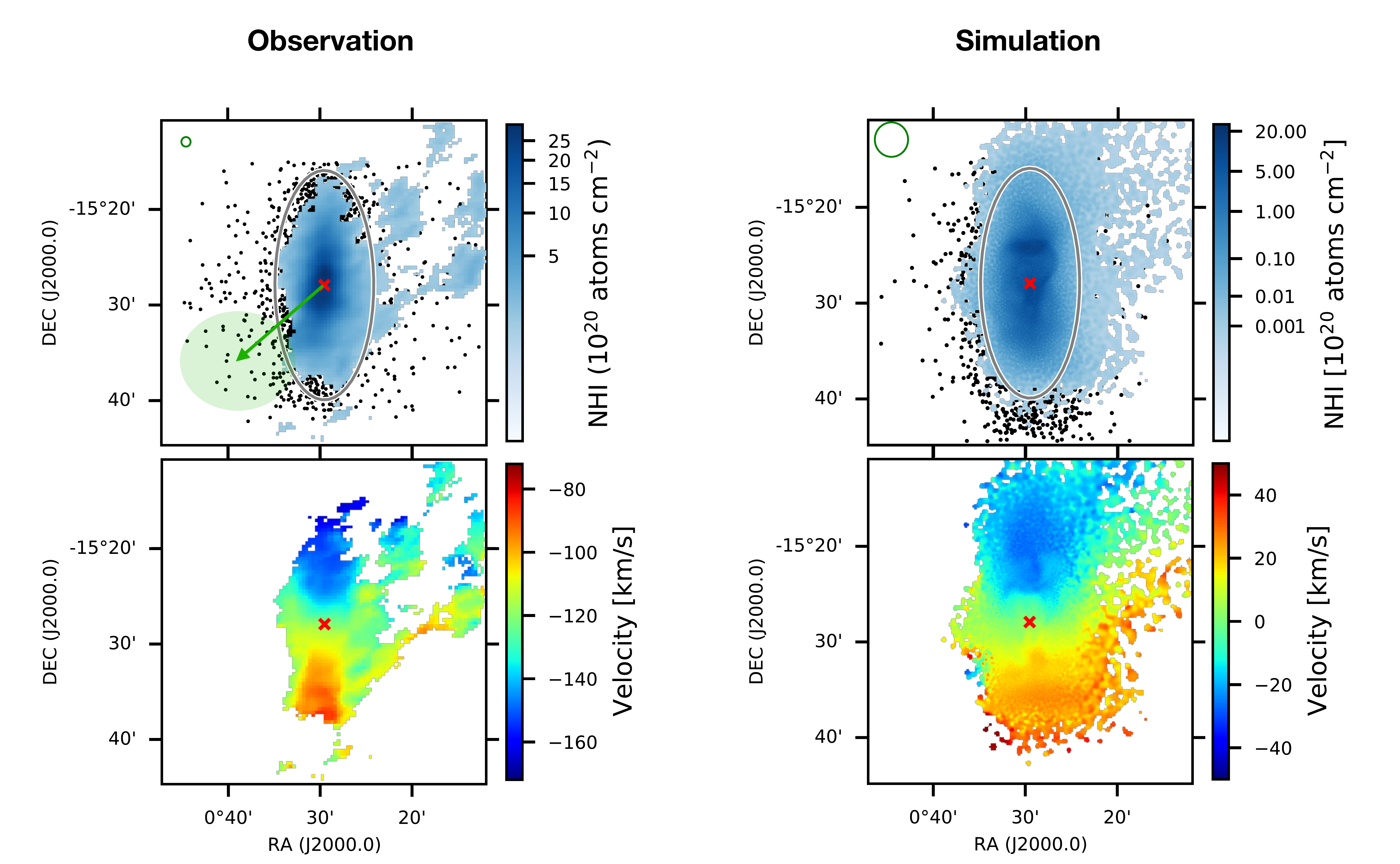}
\caption{Evidence of ram-pressure in WLM. {\it Top-left:} MeerKAT \HI\ column density map (blue) superimposed on WLM member stars (black dots) identified from the Subaru observations. The red cross indicates the optical center of WLM, and the 12$^\prime$ ellipse encloses the main body of the WLM \HI\ (the same ellipse applies to the simulation). The green arrow shows the direction of WLM's proper motion, and its one-sigma error is indicated by a transparent green circle. The small green circle  at the top-left corner indicates the beam size of the \HI\ data.  {\it Bottom-left:} MeerKAT \HI\ velocity map. {\it Right:} Same mappings but for a simulation that uses a space velocity of 500~km~s$^{-1}$ for WLM and an IGM density of $n_{\textsc{igm}} = 4 \times 10^{-6}~{\rm atoms\ cm^{-3}}$. The green circle in the {\it top-right} panel indicates the typical spatial resolution of gas particles in the simulation, i.e., twice the smoothing length of gas particles, specifically for the region near and outside the ellipse of the main body.} 
\label{fig:ram}
\end{figure*}

\section{Evidence of ram pressure in WLM}
Prompted by its anomalous rotational properties \citep{Kepley2007}, we reexamined the neutral hydrogen (\HI) properties of WLM observed by the MeerKAT radio telescope (see Appendix~\ref{sec:HI}), focusing especially on the detection of fainter features in the outskirts. We discovered four significant, extended, \HI\ clouds lying 10 to 20 arcminutes northwest of the galaxy (which corresponds to 2.7 to 5.5 kpc; see Fig.~\ref{fig:ram}). The detection of the four clouds is confirmed with signal-to-noise ratios (S/N) ranging from 20 to 28. 
We examined possible contamination by the Magellanic Stream (MS), which lies less than one degree south from WLM, using the Parkes Galactic All-Sky Survey \citep[GASS;][]{McClure-Griffiths2009,Kalberla2010} and following its in-depth analysis by \citet{Hammer2015}. Possible contamination from the MS \citep{Westmeier2008} can only affect velocities lower than $-150$~km~$\rm{s}^{-1}$ (i.e., only a small fraction of the WLM velocity range), and spatially, the \HI\ clouds lie in the opposite direction from the MS (see Fig.~\ref{fig:hi_raw}), such that a clear distinction is possible.

The amount of \HI\ gas released in the WLM outskirts represents $(6.13 \pm 0.06) \times 10^{6}M_{\odot}$, which is $\sim$10\% of the main galaxy content, $(61.8 \pm 0.03) \times 10^{6}M_{\odot}$; it is defined within a 12-arcminute ellipse in Fig.~\ref{fig:ram}. We note that a possible fraction of \HI\ material in the outskirts of WLM  may have escaped our detection as it could be outside the MeerKAT-observed field-of-view (see Fig.~\ref{fig:tail}). The two clouds closest to WLM show a bridge with the main galaxy, which indicates their physical connection to WLM. Since WLM is known to be a very isolated galaxy \citep{Higgs2016,McConnachie2021}, no interaction with another galaxy is expected.
This isolation leaves two likely possibilities for the \HI\ clouds' origin:
either a merging event near completion or ram-pressure effects. Both mechanisms could be consistent with a bridge between the \HI\ clouds and the main galaxy, but they can be distinguished by comparing the \HI\ and the optical stellar data.

We retrieved exquisite and deep Subaru Suprime-Cam imagery \citep{Miyazaki2002}, with total exposures of 36, 50, and 72 minutes in B, V, and I, respectively, and an image quality of $\sim$1.0 arcsecond. At this spatial resolution, all WLM stars are resolved, but those in WLM's core are affected by crowding.  Images were reduced, sky-subtracted, and then co-added using the Suprime-Cam Data Reduction Software \citep{Yagi2002,Ouchi2004}. We built color-magnitude diagrams (CMDs) to select WLM red giant branch (RGB) stars and eliminate contamination from the Milky Way halo stars (see Appendix \ref{sec:subaru}). No stars are detected within the four \HI\ clouds, which rules out the near-completion merger hypothesis. Further, Fig.~\ref{fig:ram} shows a precise superposition of stars with \HI\ gas (see Appendix~\ref{sec:subaru} for the precision of the astrometry), suggesting that ram pressure is responsible for the \HI\ offset in the same direction as the \HI\ clouds.

Despite its considerable distance, WLM possesses a sufficient number of stars detected by Gaia to estimate its proper motion both in direction and magnitude \citep{McConnachie2021}. The most recent Gaia Early Data Release 3 (EDR3) results \citep{Battaglia2021} provide a proper motion of $\mu^*_{\rm RA}$= 0.090$\pm$0.03 and $\mu_{\rm DEC}$= $-0.07\pm0.02$ mas yr$^{-1}$, which is consistent with the result obtained by one of us (AWM), $\mu^*_{\rm RA}$= 0.077$\pm$0.04 and $\mu_{\rm DEC}$= $-0.08\pm0.04$ mas yr$^{-1}$. We note that the Gaia systematic in proper motion, 0.03 mas yr$^{-1}$ \citep{Lindegren2021}, was not included in either of the results. This new proper motion corresponds to a total space velocity of about $300\pm150$ km/s with respect to the Milky Way center if we adopt a heliocentric radial velocity of $-130\pm 1$~km/s for WLM \citep{Leaman2009}. Since both results converge to the same value, we consider them sufficiently compelling for verifying the ram-pressure hypothesis. Figure~\ref{fig:ram} (see arrow) shows unambiguously that WLM's motion is consistent with the direction of the offset between \HI\ and stars as well as the four \HI\ clouds assuming they were stripped from WLM during its trajectory inside an external gaseous medium. 

\section{Verification by simulations}
To test the hypothesis that WLM is undergoing a ram-pressure-stripping process, we additionally performed  N-body hydrodynamical simulations based on Gizmo \citep{Hopkins2015}, a hydrodynamical solver that has been used to accurately reproduce the Magellanic Clouds' extended gas and the MS \citep{Wang2019}. We carried out a hundred or so simulations that cover a reasonably wide range of the relevant physical conditions, especially concerning the  intergalactic medium (IGM)  density and the dark matter in WLM. The details of the simulations can be found in Appendix \ref{sec:simus}.  The right panels of Fig.~1 show that it is indeed possible to reproduce \HI\ maps that closely match the observations under the assumption of ram-pressure stripping. 

\section{Discussion}
The distances between WLM and both M31 and the Milky Way are several times larger than the galaxies' nominal halo ``virial'' radii ($\le$ 300 kpc); thus, the medium responsible for the stripping is not that which we would traditionally associate with being concentrated around a normal large galaxy, unless this concentration has a remarkably large-scale radius. However, within the Local Group, the Pegasus dwarf irregular (Peg-DIG) has also been identified as a possible case of ram-pressure stripping \citep{McConnachie2007}, although this has been disputed \citep{Kniazev2009}. Regardless, observing ram-pressure stripping in Peg-DIG is still less surprising given its closer distance to M31 (475 kpc) and smaller mass than WLM. 

The threshold condition for ram pressure can be written simply as \citep{Gunn1972}
\begin{equation}
\label{eq1}
\rho_{\textsc{igm}}v^2 \geq 2 \pi G \Sigma_{\rm tot} \Sigma_g ,
\end{equation} 
\noindent where the left-hand side of the equation represents the pressure exerted by the IGM with density $\rho_{\textsc{igm}} = n_{\textsc{igm}}\mu$ on a galaxy moving at a relative velocity, $v$, and the right-hand side of the equation represents the restoring force of a disk with total surface density $\Sigma_{\rm tot}$ and gaseous interstellar medium (ISM) surface density $\Sigma_g$. We take the mean particle mass $\mu=0.75m_{\rm p}$ for fully ionized media.

The effects of ram pressure on the ISM of late-type spirals in clusters by the IGM is well illustrated by the gas-rich dwarf falling in the M81 group, Holmberg II \citep{Bureau2002}. In this case, the IGM volume density required to strip its ISM was $n_{\textsc{igm}} \geq 4.0 \times 10^{-6}\,{\rm atoms}\,{\rm cm}^{-3}$. Peg-DIG requires an environment with density $n_{\textsc{igm}} \sim\,10^{-5}$ - $10^{-6}~{\rm atoms\ cm^{-3}}$ \citep{McConnachie2007}. 
Following \citet{McConnachie2007} and \citet{Bureau2002}, we applied Eq.~(\ref{eq1}) to WLM, for which the total mass  corresponds to the baryonic mass \citep[see explanation in][]{Bureau2002}: $\Sigma_{\rm tot}=M_{\rm tot}/(\pi R^2),$ where $R=3.24$~kpc; we find $n_{\textsc{igm}} \geq 5.2 \times 10^{-5}$~cm$^{-3}$. This limit is based on the velocity of WLM relative to the Milky Way, $v=300\pm150$~km~s$^{-1}$, for which we used $v \leq 450$~km~s$^{-1}$. It might be discussed how one can account for the relative velocity of the Milky Way versus the IGM, but we think that 450 km s$^{-1}$ is already a large value for the spatial motion of a galaxy in a sparse group. The low limit for the IGM density is much higher than that of Holmberg II or that of Peg-DIG, which came as a surprise because WLM is much farther from a large galaxy.\ This further highlights the importance of the discovery that WLM is undergoing ram-pressure stripping.

In keeping with the analytical approximation in Eq.~1, we also find that our numerical models only provide good matches to the observations when we adopt very low total masses for WLM, very high densities for the IGM, and/or high space velocities for WLM. Details of these numerical findings will be presented elsewhere. However, these basic analyses point to a notably large density for the stripping medium, which is possibly suggestive of a previously unrecognized large total reservoir of mass in the Local Group or suggests that WLM is passing through a very clumpy medium, which is still unverified even by the techniques used in Cosmicflow-3 \citep{Tully2019}.
Within 2~Mpc centered on WLM, the nearest galaxy slightly more massive than WLM is IC 1613, which is a dwarf irregular galaxy (dIrr) at 380 kpc; farther away there are M31 at 820 kpc, the Milky Way at 930 kpc, and other dwarf galaxies. Using the proper motion measured from Gaia, one can infer that WLM was perhaps located near the Local Void \citep{Tully2019,Karachentsev2004} and is currently passing through the Local Sheet structure. Such properties make WLM an archetype galaxy that could be used to probe the properties of the Local Volume. Alternatively, it is possible that the previous mass estimates for WLM are incorrect because of the asymmetric rotation curve, and certainly these new findings suggest the need for revisiting the mass models of WLM given the perturbed kinematics and the previously unrecognized effects of ram-pressure stripping on the gas dynamics.

The clear signature of ram-pressure stripping visible in WLM implies that, despite its archetypical status as an isolated field dwarf, WLM is undergoing evolutionary processes more normally associated with satellite dwarf galaxies. The very presence of a medium with a sufficient density this far from either the Milky Way or M31 that is capable of stripping a galaxy with the mass of WLM is unexpected. This discovery warrants careful observation of other so-called isolated dwarfs in the nearby Universe to determine if such a signature is a universal feature when the dwarfs are observed to sufficiently low column densities. Certainly, it implies that processes that are normally associated with the suppression of star formation in satellite systems  also operate in considerably `emptier' regions of the Universe because -- importantly -- these `empty' regions are not so empty.

\section*{Acknowledgments}
We are grateful to Nobuo Arimoto and Sakurako Okamoto for their kind helps in acquiring 
the archive data of Suprime-Cam from the Subaru database. 
This work was partly based on the data obtained by Suprime-Cam at Subaru Telescope.
Simulations in this  work  were performed at the  High-performent calculation (HPC)  resources MesoPSL financed by the project Equip@Meso (reference ANR-10-EQPX-
29-01) of the program "Investissements d'Avenir" supervised by the
'Agence Nationale de la Recherche'. 
We are grateful to Phil Hopkins who kindly
shared with us the access to the Gizmo code.
The MeerKAT telescope is operated by the South African
Radio Astronomy Observatory (SARAO), which is a facility of the 
National Research Foundation, an agency of the Department
of Science and Innovation (www.sarao.ac.za).
RI acknowledges the financial support from the State Agency for Research of the Spanish Ministry of Science, Innovation and Universities through the "Center of Excellence Severo Ochoa" awarded to the Instituto de Astrof\'{\i}sica de Andaluc\'{\i}a (SEV-2017-0709), from the Consejer\'{\i}a de Transformaci$\rm\acute{o}$n Econ$\rm\acute{o}$mica, Industria, Conocimiento y Universidades de la Junta de Andaluc\'{\i}a and the European Regional Development Fund from the European Union through the grant IAA4SKA (Ref. R18-RT-3082), and the grant RTI2018-096228-B-C31 (MCIU/AEI/FEDER,UE).

F.H.\ leads the current collaboration. As the main collaborator, C.C.\ plays a very important role in all aspects of the collaboration, including all HI data analysis, text revising. F.H.\ wrote the first version of the main text of the paper, also the data reduction of the Subaru Suprime-Cam observations. A.W.M. calculated the proper motion of WLM from Gaia EDR3, and significantly revised the main text to its present version, which considerably consolidates the paper's arguments and its scientific impact. R.I.\ and G.J.\ provided the main data analysis and mapping of the MeerKAT HI observations. B.N.\ and C.C\ provide the calculation of HI mass and error estimations. C.H.\ provides the photometric calibration of the Subaru data and highlights the connection between proper motion and HI offset, and firstly point out the possibility of ram pressure to explain the observed HI clouds. Y.Y.\ does the analysis of the Subaru data from establishing the catalog until the WLM stars selection, and points out the evidence of ram pressure by comparing precisely the optical and HI data. F.H.\ and Y.Y.\ conduct the simulation configurations, and Y.Y.\ is responsible for all the technical realization and the analysis of the simulations. All authors reviewed the manuscript. 

We provide the catalog of member candidates of WLM as shown in Fig.~1, the mappings of both observation and simulation in Fig.~1, and the simulation data of the snapshot. The data-set is available publicly at https://doi.org/10.5281/zenodo.5604890. Other data are available upon request.

\section*{ORCID for corresponding authors}
Y.Y.\ https://orcid.org/0000-0001-7949-3407\\
R.I.\ https://orcid.org/0000-0003-2476-3072\\
C.C.\ https://orcid.org/0000-0001-9089-6151\\
G.J.\ https://orcid.org/0000-0003-0608-6258\\
C.H.\ https://orcid.org/0000-0001-8650-9665\\
B.N.\ https://orcid.org/0000-0003-1032-8889\\
F.H.\ https://orcid.org/0000-0002-2165-5044 \\
A.W.M.\ https://orcid.org/0000-0003-4666-6564

\bibliographystyle{aa}
\bibliography{WLMbib}

\begin{appendix}

\section{HI properties of WLM}
\label{sec:HI}
\begin{figure}[ht]
\centering
\includegraphics[width=8cm]{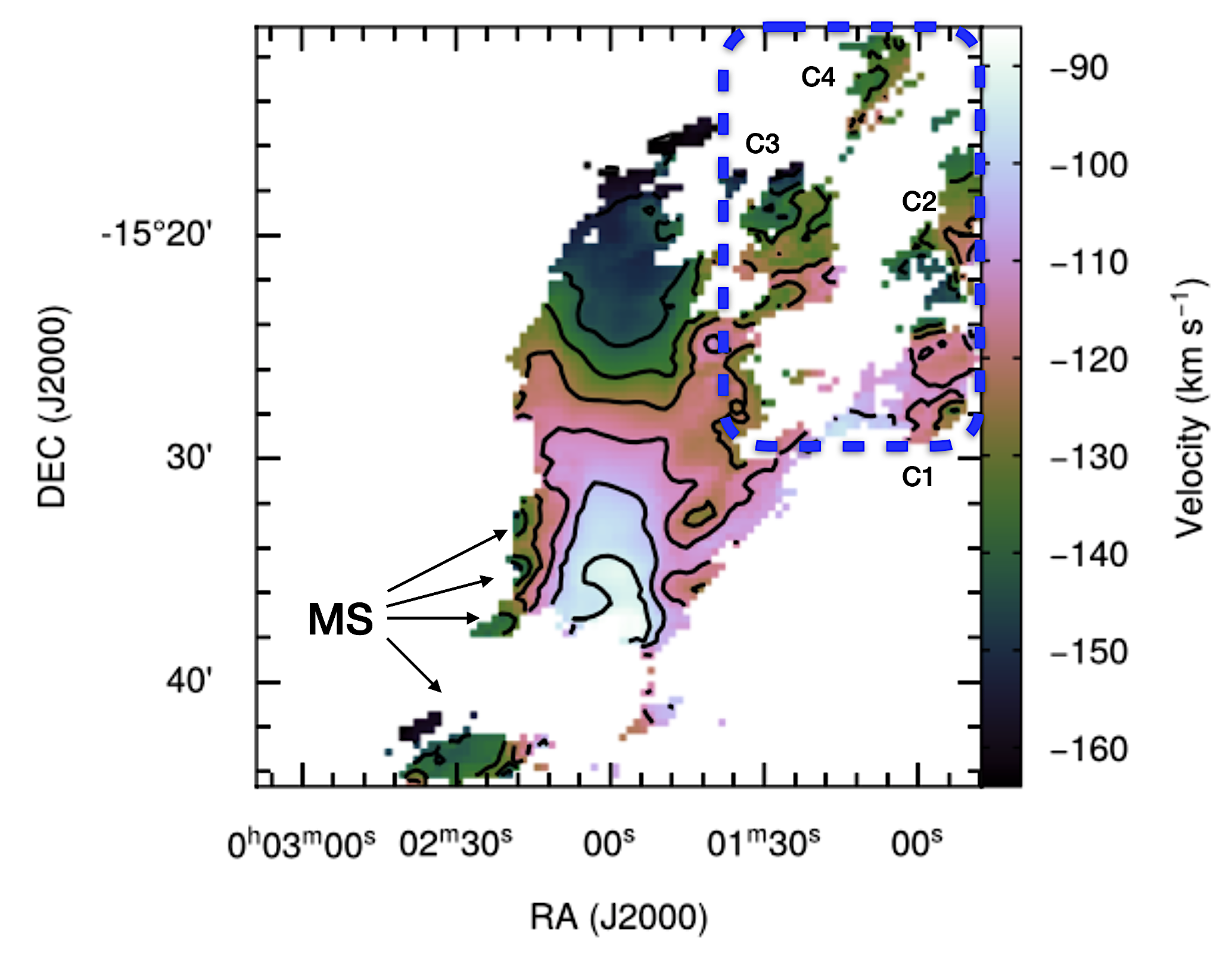}
\caption{Velocity field of WLM from the MeerKAT observations. The four discovered clouds, i.e., C1, C2, C3, and C4, and the contamination from the MS are indicated. The blue dashed-line box encloses the four clouds and also indicates the area where we measure the stripped HI gas in our simulations.
}
\label{fig:hi_raw}
\end{figure}

We probed the outskirts of the galaxy using the  MeerKAT \HI\ data published in \citet{Ianjamasimanana2020}. The \HI\ data cube covers a field-of-view of $34^\prime \times 34^\prime$ and was built with a Gaussian beam size of 60$^{\prime\prime}$ (full width at half maximum) and a velocity resolution of $5.5~\rm km~s^{-1}$ per channel \citep{Ianjamasimanana2020}. The data cube has a root mean square (RMS) noise per channel of 5 $\mathrm{mJy~beam^{-1}}$. Figure~\ref{fig:hi_raw} shows a velocity map of this data cube, which was obtained via the following analysis: We ran the MIRIAD\footnote{https://www.atnf.csiro.au/computing/software/miriad/} task MAFIA using an input mask obtained by blanking pixels in the 60$^{\prime\prime}$ cube where their flux values are below three times the RMS noise level in a lower resolution version of the cube. We then used the masked cube, which contains only genuine \HI\ emission, to derive moment maps and a velocity field. To derive the velocity field in Fig.~\ref{fig:hi_raw}, we fit each line-of-sight profile with a single Gaussian and extracted the core velocity. In the southeast of WLM, one can notice some relatively blue-shifted clouds, which show a sharp contrast in terms of kinematics ($\Delta v > 40$~km~s$^{-1}$) with respect to the southern part of WLM. After careful examinations of the Parkes GASS, we suggest that they are likely contamination from the MS  (\citealt{McClure-Griffiths2009,Kalberla2010}, see also the in-depth analysis by \citealt{Hammer2015}). Thus, we masked this contamination in our final map, which is shown in Fig.~\ref{fig:ram}.

\begin{figure}
\includegraphics[width=8cm]{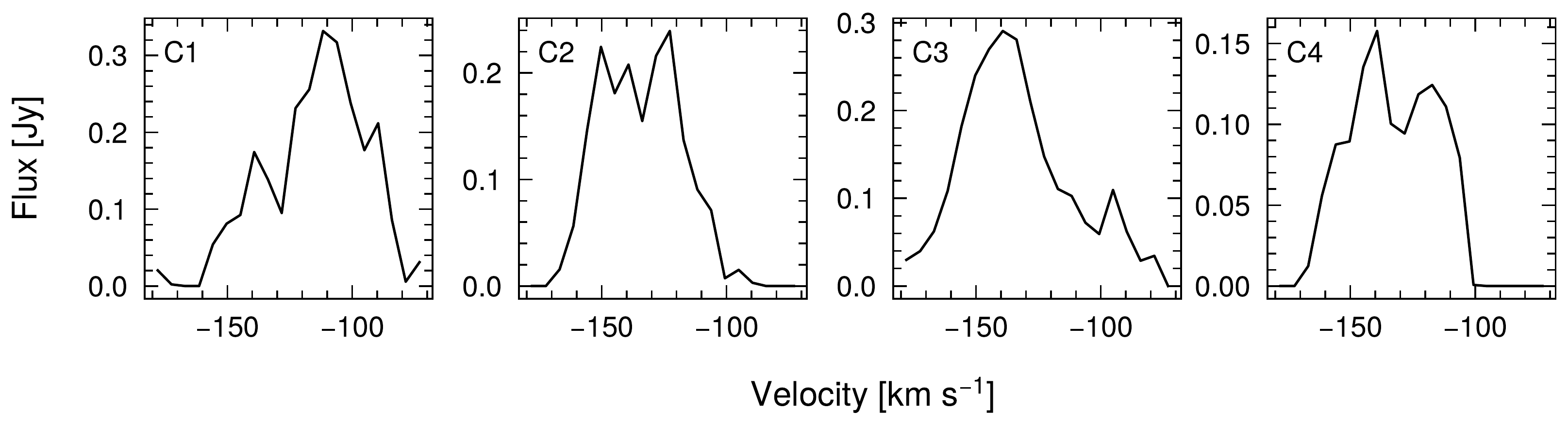}
 \caption{Integrated spectrum of the four clouds in Fig.~\ref{fig:hi_raw}.
 }
 \label{fig:wlm-mom1-final-spectrum}
\end{figure}

\begin{figure}
\centering
\includegraphics[width=8cm]{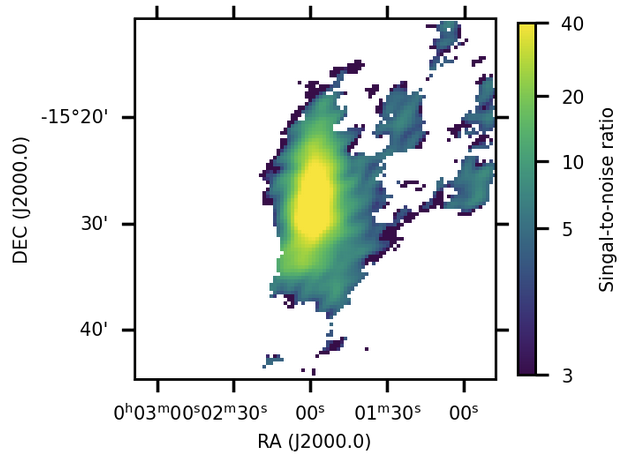}
\caption{Signal-to-noise ratio map of the MeerKAT \HI\ observations. }
\label{fig:num_datapoints}
\end{figure}

In Fig.~\ref{fig:hi_raw} we label four clouds, C1, C2, C3, and C4, and their integrated spectra are shown in Fig.~\ref{fig:wlm-mom1-final-spectrum}.  Assuming a distance of $934\pm 21$ kpc for WLM \citep{Higgs2021}, the \HI\ mass of the main body is $(61.8 \pm 0.03) \times 10^{6}M_{\odot}$. For the \HI\ clouds (from C1 to C4), we calculate $(1.93 \pm 0.09) \times 10^{6}M_{\odot}$, $(1.28 \pm 0.06) \times 10^{6}M_{\odot}$, $(2.08 \pm 0.10) \times 10^{6}M_{\odot}$, and $(0.84 \pm 0.03) \times 10^{6}M_{\odot}$, respectively. We note that for a larger distance, for example $991 \pm 23$ \citep{Lee2021}, the HI masses would be 13\% larger.
These clouds are located in the WLM velocity range and are smoothly linked into the main body of WLM on sky projection. Thus, it is likely that these clouds are indeed part of WLM. 
We assessed the reliability of the pixels belonging to these clouds by calculating the S/N of each \HI\ profile, which are shown in Fig.~\ref{fig:num_datapoints}. 

\section{Subaru observations of WLM}
\label{sec:subaru}
\begin{figure}
\centering
\includegraphics[width=8cm]{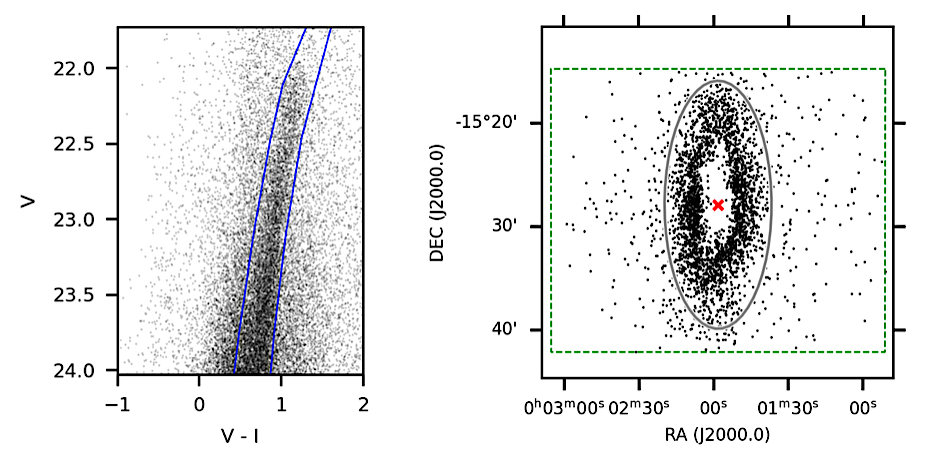}
\caption{RGB member candidates of WLM. {\it Left:} Color-magnitude diagram, V$-$I versus V. Blue lines enclose the RGB stars of WLM. {\it Right:} Space distribution of WLM member star candidates. The green dashed-line box indicates the field-of-view of the final stacked Subaru observations. The red cross and the gray ellipse indicate the optical center and the main body of WLM as in Fig.~\ref{fig:ram}. 
}
\label{fig:CMD}
\end{figure}

\begin{figure}
\centering
\includegraphics[width=8cm]{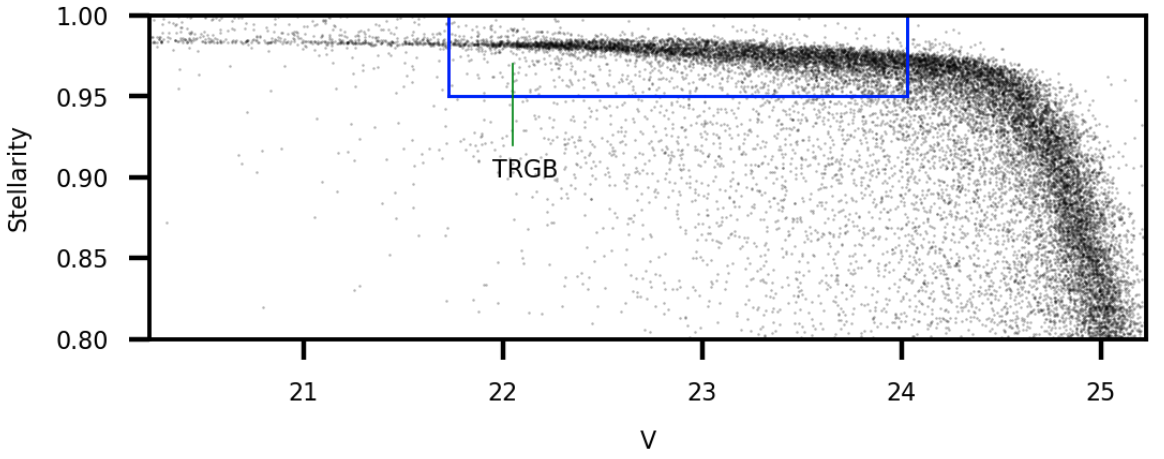}
\caption{Stellarity of sources as a function of the V-band magnitude. The blue lines indicate our selection of stellar objects and the corresponding cutoff magnitude. The green line indicates the magnitude of TRGB, i.e., 22.05 in the V band.
}
\label{fig:stellarity}
\end{figure}

We retrieved deep imagery taken with the Subaru Suprime-Cam. The observations were carried out on 2006 November 20 and covered a field-of-view of 33.5 by 27.5 arcmin$^2$ (pixel scale 0.20$^{\prime\prime}$/pixel) centered on WLM. Raw images were processed (including sky subtraction and stacking) using the Subaru Suprime-Cam standard pipeline \citep{Yagi2002,Ouchi2004}\footnote{https://www.naoj.org/Observing/Instruments/SCam/sdfred/sdfred1.html.en} in the World Coordinate System frame. We obtained the final stacked images in the B, V, and I bands\footnote{Specifically, W-J-B, W-J-V, and W-J-IC, respectively.} with total exposure times of 36, 51, and 72 minutes and seeings (full width at half maximum of the Gaussian fit) of 1.05, 0.82, and 1.07 arcsec, respectively.
With SExtractor \citep{Bertin1996}, we first detected all sources three sigma above the background in each band. The ``LOCAL'' background and de-blending options were adopted for source detection. Taking the V band as the reference, we cross-identified all the sources detected simultaneously in the three bands and generated the final raw source catalog. We adopted the default artificial neural networks technique in SExtractor for star-galaxy morphology classification, characterized by the {\tt stellarity} parameter (approaching ``1'' for point sources and approaching ``0'' for extended sources.) Due to the central crowding, it is very difficult to properly recover the sources there (hence the ``hole'' in the center in Fig.~\ref{fig:CMD}). Our real interest is to detect WLM stars in its outskirts in particular, to verify whether the \HI\ clouds in the WLM outskirts have optical counterparts. To calibrate the photometry, we used a simple, first-order comparison to the existing WLM observations from the Solo Survey \citep{Higgs2016,Higgs2021} as photometry is not critical to the selection of RGB stars. The previous observations of WLM were converted to the V and I bands using the technique provided in \citet{Thomas2021}. The tip of the red giant branch (TRGB) was identified using a five-point Sobel filter in both the Solo and Subaru observations. The median color of RGB stars within 1 magnitude of the TRGB was identified. Using the TRGB and the median color, we obtained a simple calibration for the Subaru CMD. 
We selected RGB member candidates using the combination of the CMD of V versus V$-$I (Fig.~\ref{fig:CMD}) plus ${\tt stellarity}>0.95$ (see Fig.~\ref{fig:stellarity}) and a magnitude cutoff of $21.73<$V$ < 24.03$ (where the S/N for individual sources is larger than 18).  Finally, 3341 stars were obtained as the best member candidates of WLM (black dots in Figs.~\ref{fig:ram} and \ref{fig:CMD}). In order to compare directly with the HI map, the astrometric position of the center of WLM was calibrated using the Gaia EDR3 catalog in the field-of-view.  We evaluated the mean position of 103 sources (around the main body region of WLM) that are cross-identified in the Gaia EDR3 catalog. We found mean shifts of $2.6$ and $5.4$ arcseconds in right ascension and declination, respectively, and these shifts have been corrected in Figs.~\ref{fig:ram} and \ref{fig:CMD}.

\section{N-body hydrodynamical simulations}
\label{sec:simus}
To illustrate the effect of ram pressure acting on WLM, we ran simulations to sample the possible range of parameters. For the galaxy initial condition (IC), we assumed a stellar mass of $1 \times10^{7}M_\odot$ \citep{Lelli2016} and an \HI\ mass of $1 \times10^{8}M_\odot$ \citep{Kepley2007}, considering some of the gas will be stripped by ram pressure. The galaxy IC is composed of a thick stellar disk, a gas disk, and a dark matter halo following the method from \citet{Barnes2002}. We considered two types of ICs to represent WLM after considering its asymmetric rotation curve \citep{Ianjamasimanana2020, Kepley2007}, WLM7 and WLM8, which have maximal rotation velocities of about 30 and 40 km/s, respectively.  We adopted a mass resolution of 1375 $M_\odot$ per particle for both the stellar and gas components in the dwarf galaxy and 5512.4 $M_\odot$ per dark matter particle. 

\begin{figure}
\centering
\includegraphics[width=9cm]{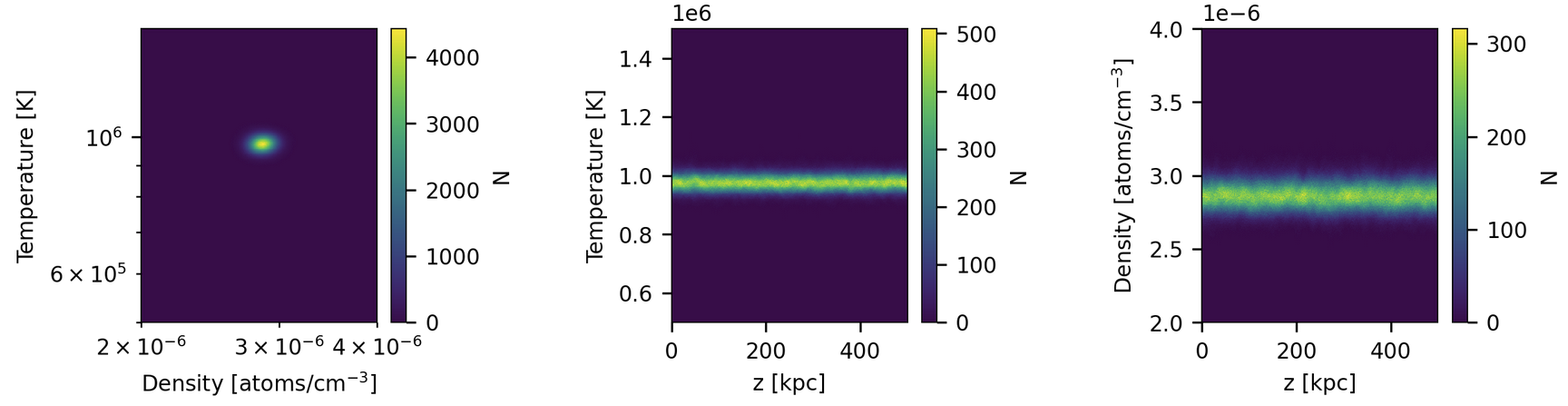}
\caption{Phase diagrams of the IGM cube with ($\rho_{\textsc{igm}}=2.9 \times 10^{-6}$~atoms~cm$^{-3}$) after 2.5 Gyr in isolation.
}
\label{fig:igmcube}
\end{figure}

Since WLM is located at the edge of the Local Group, almost 1 Mpc from both the Milky Way and M31, we neglected the gravity from both galaxies. In our physical model WLM is traveling through the IGM at a constant velocity. To simulate the IGM, we first set up a 500-kiloparsec-wide cube filled with a uniform density of gas at a mean temperature of 1 million kelvin. We chose a mass resolution of 1375 $M_\odot$ per particle to simulate the IGM, the same as for the galaxy cold gas. We used the ``periodical'' condition in Gizmo to run the cube, meaning each surface of the cube is eventually joined with the surface on the opposite side of the cube in a round way. This technique provides a very good stability in terms of thermal properties, as well as gravity inside the cube. 
Figure.~\ref{fig:igmcube} shows a phase diagram of this cube, which can stay unchanged for a few billion years when isolated. 
If a galaxy is moving toward a given external surface of the cube, the galaxy will pass through the surface and then appear in the cube from the opposite side thanks to the periodical condition. 
We only need a tunnel when we study the ram-pressure effects, and thus we took advantage of the cube and designed 16 tunnels, each with a volume of $125\times 125\times 500$ kpc$^3$. 
We can run 16 galaxies in parallel in the cube. This is a very efficient way to sample the parameter spaces -- especially for changing parameters, such as the inclination of the galaxy with respect to its motion direction -- and then observe the ram-pressure forces.
The maximal velocity tested in our simulation is 500 km/s, meaning a galaxy takes about 1 Gyr to pass through one cube side to another. 
 To avoid the wake of ram-pressure tails that may affect the neighboring tunnels, we required all 16 galaxies to have the same velocity in the given cube. This ensures the same IGM conditions during their motions. In fact, due to the high sound speed (155 km/s) of the IGM, the average properties of the IGM recover their initial levels in a few hundred million years after the galaxy passages.
Before putting the galaxy ICs into the cube, we ran them in isolation for 2 Gyr to completely stabilize them. All simulations were performed with Gizmo, including the physical module of gas cooling and star formation \citep{Wang2012,Cox2006}, and with the advanced technique of adaptive gravitational softening \citep[AGS;][]{Hopkins2015,Price2007}. When applying AGS, we adopted minimal softenings of 0.03 and 0.04 kpc for the baryonic and dark matter particles, respectively, knowing that we set the dark matter particles to be four times more massive than the baryon particles. \\

\HI\ observations show that the direction of the gas striping is 45 degrees West of North (see Fig.~\ref{fig:ram}). We adopted the WLM inclination value of 75 degree from \citet{Leaman2012}. These two angles can be easily converted and fixed into the simulation frame with respect to the tunnel axis, which then follows the galaxy motion. We sampled two possible inclinations of WLM because we do not know which side, the left or the right, is the closest to us. However, the WLM motion direction with respect to the IGM is perhaps the least constrained parameter. From the proper motion and line-of-sight velocity we can infer the WLM velocity relative to the Milky Way, but we know how neither the Milky Way itself nor WLM moves in the IGM. Part of this freedom, defined by $\Theta$, corresponds to the component of the gas striping direction projected on our line-of-sight. We sampled all possible values of $\Theta$ by steps of 20 degrees and narrowed it down to a range of around $-$10 to 30 degrees,  which seems sufficient for our illustration purposes. The case of $\Theta=0$ corresponds to the mean ram-pressure force acting along the same axis of the WLM space motion relative to the mass center of the Milky Way. \\

To simulate WLM in the IGM, we let each galaxy run for 2 Gyr in the tunnel. We skipped the analysis of the first 0.5 Gyr to avoid any numerical artifacts because we manually put a galaxy with a cold gas disk directly into the hot medium, although we noticed that the system could have already been stabilized in the first 0.2 Gyr. Simulated galaxies were analyzed in the observational frame, by aligning the stellar mass center of the simulated galaxy at the WLM stellar mass center and at a distance of 932 kpc \citep{McConnachie2005}. Only 60-kiloparsec cubes around each galaxy were analyzed. The stellar particles are directly plotted and are compared to the observed stellar distribution. The \HI\ column density map was evaluated through the following algorithm.  First, we identified the cold gas (i.e., the \HI\ gas) as the gas with temperature $T < 20000$~K and assumed that each gas particle occupies a cubic volume. The size of the volume can be calculated using $l = \sqrt[3]{m/\rho}$, where $m$ and $\rho$ are the mass and the density of each gas particle. By projecting the volume onto the sky, we obtain the average N$_{\footnotesize\textsc{H\,i}}$ = $X\rho\,l$ on a surface $l^2$, where $X$ is 0.76 the hydrogen fraction.  For a better visualization, we redistributed the total N$_{\footnotesize\textsc{H\,i}}$ in the surface using a Gaussian kernel with a $\sigma = l/6$. The factor 6 is an arbitrary choice for a better visualization.
By stacking the N$_{\footnotesize\textsc{H\,i}}$ column density of all gas particles, we obtained the N$_{\footnotesize\textsc{H\,i}}$ map of simulated galaxies.
This algorithm can be easily applied to obtain a data cube by adding the third dimension, $-v_z$, of the simulation, which corresponds to the observed line-of-sight velocity. For each spaxel of the data cube, we calculated the N$_{\footnotesize\textsc{H\,i}}$-weighted average velocity, providing the velocity map shown in Fig.~\ref{fig:ram}. A mask that limits N$_{\footnotesize\textsc{H\,i}}>10^{16}$~atoms/cm$^2$ has been applied to both the N$_{\footnotesize\textsc{H\,i}}$ maps and the velocity map.

\begin{figure}
\centering
\includegraphics[width=8cm]{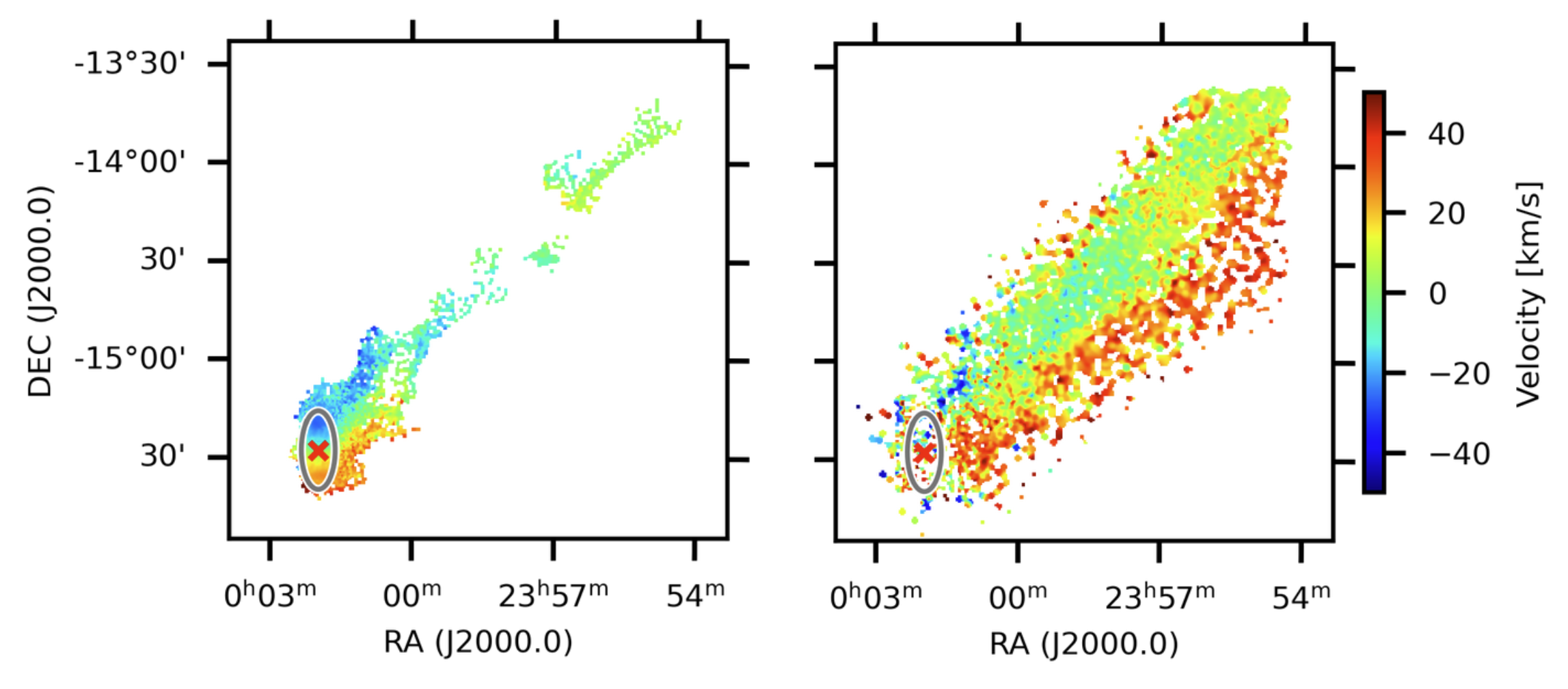}
\caption{Overview of the velocity field of the stripped gas for the simulation presented in Fig.~\ref{fig:ram} at epoch 1.7~Gyr for \HI \ ({\it left}) and for \textsc{H ii} ({\it right}).  A threshold of N$_{\textsc{H\,i}}$ at $10^{15}$~atoms~cm$^{-2}$ is applied for both. Since we only analyzed a 60-kiloparsec-wide cube around the galaxy, we see a sharp cutoff of the data in the \textsc{H\,ii} map. The galaxy is located at the left bottom of the figure, as indicated by the red cross and the gray ellipse, which are the same as in Fig.~\ref{fig:ram}.
}
\label{fig:tail}
\end{figure}

To compare with the stripped clouds in the observations, we generated a statistic of the simulation results from 68 models, including two ICs of different masses for WLM, different IGM densities varying from  $1\times10^{-6}$ to $2.9\times 10^{-5}$~atoms~cm$^{-3}$, and different space velocities from 300 to 500~km~s$^{-1}$, as well as the orientation of galaxy disk with respect to ram-pressure forces. For each model, we measured the total \HI\ mass inside the main body within an ellipse of 12 arcmin, as for the observation, and the \HI\ mass that corresponds the cloud regions (i.e., the blue box in Fig.~\ref{fig:hi_raw}). We find that the higher density of the IGM helps strip more gas from the galaxy because the residual \HI\ in the main body decreases with the increase in the IGM density. For all possible configurations, the stripped \HI\ gas is always below 3-4\%. One possible reason could be that although a higher IGM density helps strip more gas from the galaxy, it also leads to strong Kelvin-Helmholtz instabilities, which heats the larger fraction of stripped gas into the ionized phase (i.e., \textsc{H ii}), and this gas is disturbed over a very large area (Fig.~\ref{fig:tail}).

\end{appendix}

\end{document}